\documentclass[11pt,letterpaper]{article}
\usepackage{systeme} 
\usepackage[utf8]{inputenc}
\usepackage{multirow} 
\usepackage{booktabs} 
\usepackage[english]{babel}
\usepackage{amsmath}
\usepackage{amsfonts}
\usepackage{amssymb}
\usepackage{graphicx}
\usepackage[small,bf]{caption} 
\usepackage[left=3cm,right=3cm,top=2cm,bottom=3cm]{geometry}
\usepackage{framed} 
\usepackage{color} 
\usepackage{wrapfig}\definecolor{shadecolor}{RGB}{220,220,220} 
\usepackage{float} 
\usepackage{array} 
\usepackage{caption} 

\author{Jorge Pinochet}
\title{\textbf{Hawking in a thousand words}}
\begin{document}

\author{Jorge Pinochet$^{*}$\\ \\
 \small{$^{*}$\textit{Facultad de Ciencias Básicas, Departamento de Física. }}\\
  \small{\textit{Centro de Desarrollo de Investigación CEDI-UMCE,}}\\
 \small{\textit{Universidad Metropolitana de Ciencias de la Educación,}}\\
 \small{\textit{Av. José Pedro Alessandri 774, Ñuñoa, Santiago, Chile.}}\\
 \small{e-mail: jorge.pinochet@umce.cl}\\}

\date{}
\maketitle

\begin{center}\rule{0.9\textwidth}{0.1mm} \end{center}
\begin{abstract}
\noindent British physicist Stephen Hawking’s most important discovery was that “black holes are not so black,” as they possess a temperature and emit thermal radiation. In his popular science texts, Hawking offered a detailed explanation of this phenomenon. The aim of this work is to translate that explanation into mathematical language accessible to an advanced high school student, all within a thousand words.\\ \\


\begin{center}\rule{0.9\textwidth}{0.1mm} \end{center}
\end{abstract}

\maketitle

\section{Introduction}

British physicist Stephen Hawking's most important discovery was that "black holes are not so black" [1], since they emit thermal radiation as if they were hot bodies with an absolute temperature calculated using an equation known as the \textit{Hawking temperature}:

\begin{equation}
T_{H} = \frac{\hbar c^{3}}{8 \pi kGM},
\end{equation}

where $M$ is the mass of the black hole, and $\hbar$, $c$, $k$, $G$ are fundamental physical constants. The detailed derivation of this equation requires advanced knowledge of physics and mathematics. The aim of this work is to derive it and explain its physical meaning using only high school algebra, all within a thousand words. To achieve this, we will draw upon a physical picture introduced by Hawking in various popular science texts, based on the so-called \textit{quantum fluctuations of the vacuum} [1–3]. Specifically, we will translate this physical picture into mathematical language accessible to an advanced high school student. Readers interested in exploring the Hawking temperature in greater depth can consult the various educational articles published on this topic [4–11].

\section{Quantum Fluctuations and Hawking Temperature}

Before Hawking came on the scene, it was assumed that black holes do not emit any radiation, since otherwise they would not be black.

\begin{figure}[H]
  \centering
    \includegraphics[width=0.25\textwidth]{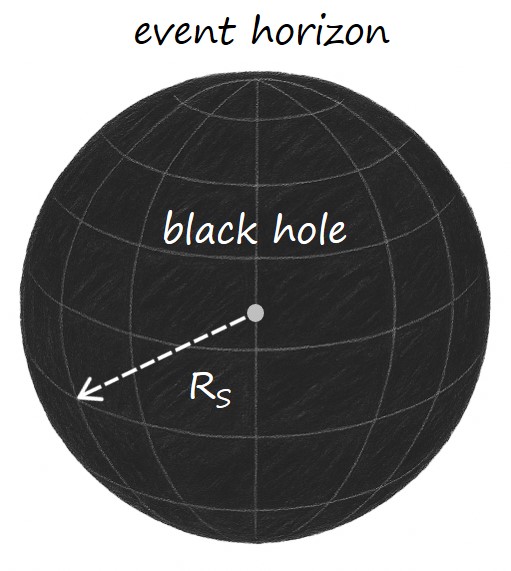}
  \caption{A static black hole has a spherical horizon of radius $R_{S}$.}
\end{figure}

However, Hawking wondered whether the effect of quantum fluctuations near the \textit{event horizon} of a static black hole could modify this scenario. Recall that the event horizon (Fig. 1) is a spherical region that defines the size of a black hole, and that in the case of a static black hole it has a characteristic radius called the \textit{Schwarzschild radius}, which is calculated as [3,12]

\begin{equation}
R_{S} = \frac{2GM}{c^{2}},
\end{equation}

where $M$ is the mass of the black hole, $G = 6.67 \times 10^{-11} N\cdot m^{2} \cdot kg^{-2}$ is the gravitational constant, and $c \cong 3\times 10^{8} m \cdot s^{-1}$ is the speed of light in vacuum.\\

\begin{figure}[H]
  \centering
    \includegraphics[width=0.3\textwidth]{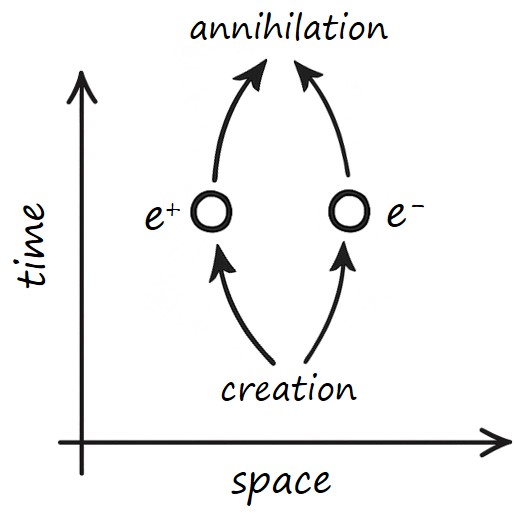}
  \caption{An electron-positron pair suddenly appears in a vacuum. The particles separate and then come together again to annihilate each other and disappear.}
\end{figure}

Quantum fluctuations are a consequence of Heisenberg's uncertainty principle, according to which the smallest value that the product of the uncertainty in energy $\Delta E$ and the uncertainty in time $\Delta t$ can take is [13,14]:

\begin{equation}
\Delta E \Delta t = \hbar /2,
\end{equation}

where $\hbar = h/2\pi = 1.05 \times 10 ^{-34} J \cdot s$ is the reduced Planck constant. This equality implies that if the vacuum energy were always exactly zero ($\Delta E=0$) then $\hbar$ would also have to be zero, which is false. Therefore, the vacuum cannot be perfect and must fluctuate, giving rise to the momentary appearance of \textit{virtual particles}, which rapidly disappear in time $\Delta t = \hbar /2\Delta E$ [9]. These particles do not violate conservation of energy, since their existence is fleeting and protected by the uncertainty relation. However, since electric charge is a conserved quantity and does not obey the uncertainty principle, virtual particles must always appear in particle-antiparticle pairs whose net charge is zero. An example is an electron-positron pair, where the particles have charges of equal magnitude but opposite sign, so they cancel each other out (Fig. 2).\\

Since $c$ is the maximum speed allowed by the laws of physics, Eq. (3) reveals that the uncertainty $\Delta x$ in the location of a pair of virtual particles cannot be less than $c\Delta t$, so that

\begin{equation}
\Delta x = c\Delta t \sim \frac{c \hbar}{E}.
\end{equation}

According to Hawking’s picture, presented in his popular science texts, quantum fluctuations near the event horizon can give rise to a process in which tidal forces separate a pair of virtual particles [1–3]. One falls into the black hole with negative energy, while the other escapes as a real particle with positive energy, thus conserving the total energy (Fig. 3). A distant observer perceives these emitted particles as thermal radiation with a temperature given by Eq. (1) (Fig. 4) [4].\\

We can translate the above description into the language of high school algebra using a simple heuristic argument. Since dimensionless numerical factors are unimportant in heuristic calculations, we will omit them from the results for simplicity.

\begin{figure}[H]
  \centering
    \includegraphics[width=0.4\textwidth]{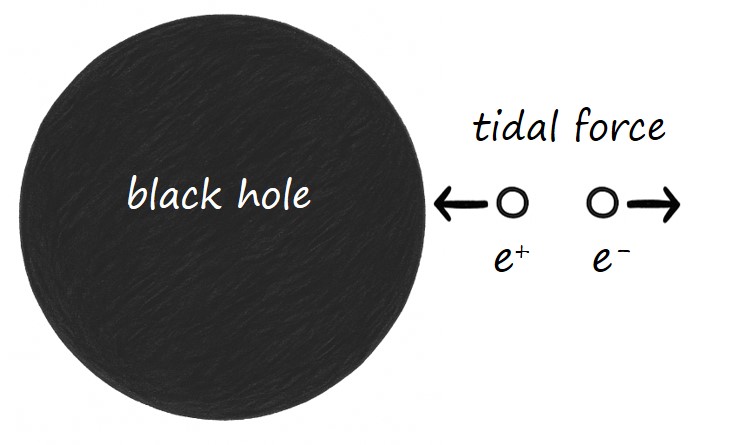}
  \caption{An electron-positron pair near the horizon is pulled apart by powerful tidal forces. The closer particle is absorbed, and the farther one is free to escape to infinity.}
\end{figure}

Let us begin by recalling that, according to Newton's law of universal gravitation, the magnitude of the tidal force $F$ exerted by a spherical celestial body of mass $M$ on two particles of mass $m$ separated by a distance $r$ from the center of the celestial body, is [15]

\begin{equation}
F= \frac{2GMml}{r^{3}} \sim \frac{GMml}{r^{3}},
\end{equation}

where $l\ll r$ is the separation between the particles. It is worth mentioning that Eq. (5) can be easily obtained by differentiating the law of universal gravitation with respect to the radius, which introduces a negative sign that we have ignored. Let us use this equation to calculate the tidal force exerted by the black hole on a pair of virtual particles. Since quantum fluctuations occur very close to the event horizon, in Eq. (5) we can take $r \sim R_{S} \sim GM/c^{2}$, so that the magnitude of the tidal force on the virtual pair is given by

\begin{equation}
F  \sim \frac{GMml}{R_{S}^{3}} \sim \frac{mc^{6}l}{G^{2}M^{2}},
\end{equation}

and the work required to separate the particles by a distance $l$ is

\begin{equation}
W \approx Fl \sim \left( \frac{mc^{6}l}{G^{2}M^{2}} \right)l = \frac{mc^{6}}{G^{2}M^{2}}l^{2}.
\end{equation}
 
By Einstein's mass-energy equivalence we have that $m \sim E/c^{2}$, and therefore,

\begin{equation}
W \sim \frac{Ec^{4}}{G^{2}M^{2}}l^{2}.
\end{equation}

\begin{figure}[H]
  \centering
    \includegraphics[width=0.45\textwidth]{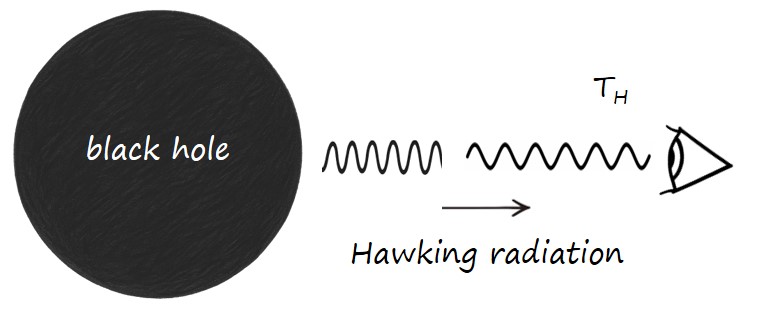}
  \caption{For a distant observer, the effect of quantum fluctuations near the horizon is perceived as thermal radiation with temperature $T_{H}$.}
\end{figure}

From Eq. (4), let us assume that the distance that a pair of virtual particles can be separated is $l \sim \hbar c/E$ and let us further assume that $W \sim E$. Under these conditions, Eq. (8) can be rewritten as $E \sim c^{6} \hbar ^{2}/G^{2} M^{2} E$, and solving this equality for $E$ we obtain

\begin{equation}
E \sim \frac{\hbar c^{3}}{GM}.
\end{equation}

But Hawking radiation particles are thermal, and according to statistical thermodynamics their typical energy should be $E \sim kT$ [16], where $T$ is the absolute temperature, and $k =1.38 \times 10^{-23} J\cdot K^{-1}$  is the Boltzmann constant, so that by Eq. (9) $T \sim \hbar c^{3}/kGM$. If we want to convert this into an equality, we must incorporate the dimensionless constants that would have appeared in an exact calculation. Grouping these constants into the dimensionless quantity $\gamma$, we have:

\begin{equation}
T_{H} = \gamma \frac{\hbar c^{3}}{kGM}.
\end{equation}

Comparing this equation with Eq. (1), we see that $\gamma =1/8 \pi$. Despite the simplicity of the derivation, this result is equal to the Hawking temperature, which was originally obtained by the British genius using much more complex mathematical arguments.\\

The Hawking temperature has been widely recognized as one of the greatest scientific discoveries of the 20th century and represents a decisive advance in our understanding of the universe.

\section*{Acknowledgments}
I would like to thank Daniela Balieiro for her valuable comments on the writing of this paper.

\section*{References}

[1] S.W. Hawking, A brief history of time, Bantam Books, New York, 1998.

\vspace{2mm}

[2] S.W. Hawking, Black Holes and Thermodynamics, Physical Review D 13 (1976) 191–197.

\vspace{2mm}

[3] S.W. Hawking, The Universe in a Nutshell, Bantam Books, New York, 2001.

\vspace{2mm}

[4] J. Pinochet, Hawking for everyone: commemorating half a century of an unfinished scientific revolution, Phys. Educ. 59 (2024) 055001. https://doi.org/10.1088/1361-6552/ad589c.

\vspace{2mm}

[5] J. Pinochet, Three easy ways to the Hawking temperature, Physics Education 56 (2021) 053001. https://doi.org/10.1088/1361-6552/AC03FC.

\vspace{2mm}

[6] J. Pinochet, Hawking for beginners: a dimensional analysis activity to perform in the classroom, Phys. Educ. 55 (2020) 045018.

\vspace{2mm}

[7] J. Pinochet, Five misconceptions about black holes, Phys. Educ. 54 (2019) 55003.

\vspace{2mm}

[8] J. Pinochet, “Black holes ain’t so black”: An introduction to the great discoveries of Stephen Hawking, Phys. Educ. 54 (2019) 035014.

\vspace{2mm}

[9] J. Pinochet, The Hawking temperature, the uncertainty principle and quantum black holes, Phys. Educ. 53 (2018) 065004.

\vspace{2mm}

[10] J. Pinochet, Hawking Temperature: An elementary approach based on Newtonian Mechanics and Quantum Theory, Phys. Educ. 51 (2016) 015010.

\vspace{2mm}

[11] M.C. LoPresto, Some Simple Black Hole Thermodynamics, The Physics Teacher 41 (2003) 299–301.

\vspace{2mm}

[12] B. Schutz, Gravity from the Ground Up, Cambridge University Press, Cambridge, 2003.

\vspace{2mm}

[13] K. Krane, Modern Physics, John Wiley and Sons, Hoboken, 2012.

\vspace{2mm}

[14] P.A. Tipler, R.A. Llewellyn, Modern Physics, W. H. Freeman and Company, New York, 2012.

\vspace{2mm}

[15] M.L. Kutner, Astronomy:A Physical Perspective, Cambridge University Press, New York, 2003.

\vspace{2mm}

[16] F. Reif, Fundamentals of Statistical and Thermal Physics, Waveland Press, Illinois, 2009.

\end{document}